\documentclass[%
 reprint,
superscriptaddress,
showpacs,preprintnumbers,
 amsmath,amssymb,
]{revtex4-1}

\usepackage{cancel}
\usepackage{mciteplus,slashed}
\usepackage{amssymb,cancel,amsmath}
\usepackage{dcolumn}
\usepackage{bm}
\usepackage{subcaption}
\usepackage{appendix}
\usepackage{feynmp-auto}
\usepackage[T1]{fontenc}	
\usepackage{csvsimple}
\usepackage{hyperref}
\usepackage[section]{placeins}
\usepackage[capitalise]{cleveref}
\usepackage{booktabs}
\newcommand{\be}{\begin{equation}}
\newcommand{\ee}{\end{equation}}
\newcommand{\ba}{\begin{align*}}
\newcommand{\ea}{\end{align*}}
\newcommand{\bpm}{\begin{pmatrix}}
\newcommand{\epm}{\end{pmatrix}}
\newcommand{\bea}{\begin{eqnarray}}
\newcommand{\eea}{\end{eqnarray}}

\newcommand{\benum}{\begin{enumerate}}
\newcommand{\eenum}{\end{enumerate}}
\newcommand{\bi}{\begin{itemize}}
\newcommand{\ei}{\end{itemize}}

\newcommand{\I}{\dot{\imath}}
\newcommand{\ra}{\rightarrow}

\newcommand{\GeV}{~\mathrm{GeV}}

\newcommand{\gsim}{\lower.7ex\hbox{$\;\stackrel{\textstyle>}{\sim}\;$}}
\newcommand{\lsim}{\lower.7ex\hbox{$\;\stackrel{\textstyle<}{\sim}\;$}}

\usepackage{graphicx}

\begin{document}

\preprint{APS/123-QED}


\title{
Neutrino Trident Production at the Intensity Frontier}

\author{Gabriel Magill}
\email{gmagill@perimeterinstitute.ca}
\author{Ryan Plestid}
\email{plestird@mcmaster.ca}
\affiliation{\mbox{Department of Physics \& Astronomy, McMaster University, 1280 Main St. W., Hamilton, Ontario, Canada}}
\affiliation{Perimeter Institute for Theoretical Physics, 31 Caroline St. N., Waterloo, Ontario, Canada}
\date{\today}

\begin{abstract}
 We have calculated cross sections for the production of lepton pairs by a neutrino incident on a nucleus using both the equivalent photon approximation, and deep inelastic formalism. We find that production of mixed flavour lepton pairs can have production cross sections as high as 35 times those of the traditional $\nu_\mu\rightarrow\nu_\mu\mu^+\mu^-$ process. Rates are estimated for the SHiP and DUNE intensity frontier experiments. We find that multiple trident production modes, some of which have never been observed, represent observable signals over the lifetime of the detectors. Our estimates indicate that the SHiP collaboration should be able to observe on the order of 300 trident events given $2\cdot 10^{20}$ POT, and that the DUNE collaboration can expect approximately 250 trident events in their near detector given $3\cdot 10^{22}$ POT. We also discuss possible applications of the neutrino trident data to be collected at SHiP and DUNE for SM and BSM physics.
\end{abstract}

\maketitle


\section{Introduction \label{sec:Intro}}
Neutrino physics has traditionally been dominated by the measurement of oscillation parameters and the study of neutrino nucleus scattering. These experimental signals are largely dominated by charged current (CC), and neutral current (NC) interactions whose cross sections scale as $\sigma\thicksim sG_F^2$. Traditionally, limits on beam luminosity have resulted in event counts that leave sub-dominant processes with expected event rates less than unity in the lifetime of an experiment. As a result these processes are often omitted in the discussions of neutrino physics. One such neglected process is neutrino trident production which has been previously observed at CHARM II, CCFR, and NuTev \cite{Geiregat1990,Mishra1991,NuTeVCollaboration1998}. 
%
%
These measurements provided evidence at the $3\sigma$ level for the contribution of Z bosons in weak interactions \cite{Mishra1991}, and more recently have been used to constrain BSM physics. Specifically, measurements from CCFR currently provide the best constraints on the mass and coupling of a heavy $Z'$ force-mediator charged under $L_\mu-L_\tau$ \cite{Altmannshofer2014}. Both of these applications are successful because the neutrino trident production of leptons is sensitive to both the vector and axial current couplings (see \cref{sec:Lepton-Matrix}). 

\begin{figure}[!h]
  \centering	\includegraphics[width=0.85\linewidth]{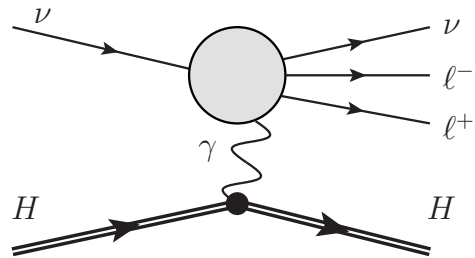}
  \caption{Leading hadronic contribution to trident production. Arrows denote direction of momentum.\
    \label{fig:schematic-trident}}
\end{figure}

The aforementioned collaborations only measured one possible mode of trident production; specifically $\nu A \rightarrow \nu \mu^+ \mu^- A$. The leading order contribution to this process involves the production of a muon-anti-muon pair, which can then interact with the target nucleus $A$ electromagnetically (see \cref{fig:schematic-trident}). For low momentum transfers ($Q\ll R_A^{-1}$) the nucleus interacts coherently with the virtual photons ($\sigma \propto Z^2$), and there is a strong enhancement due to the infrared divergence in the  photon propagator; it is this kinematic regime which dominates the cross section. Other qualitatively similar processes, such as $e^+  e^-$ or $\mu^+e^-$ trident production, were kinematically accessible, however, due to technological limitations in the detector design, the required vertex resolution for trident identification was not achievable for electrons. This would not be an issue with modern detectors. 

The cross section for $\mu^+\mu^-$ neutrino trident production is approximately five orders of magnitude smaller than the charged current cross section ($\sigma\approx 10^{-5} \sigma_{CC}$) for a $50~\text{GeV}$ neutrino scattering off an iron nucleus \cite{Belusevic1988}; high $Z$ materials will have an even larger cross section relative to CC scattering. This means that practically trident production can only be observed in experiments with very large neutrino fluxes. Additionally the leading contribution to the cross section discussed in the preceding paragraph can be calculated using the equivalent photon approximation and scales as $\sigma \thicksim G_F^2 E_\nu Q_\text{max}  \log(E_\nu Q_\text{max}/m^2_\ell)$, where  $m_\ell$ is related to the lepton masses, and $Q_\text{max}$ is a characteristic momentum transfer set by the radius of the nucleus \cite{Belusevic1988}. These considerations imply that for trident to be a useful tool one needs to consider experiments with both a high energy neutrino beam ($\langle E_\nu\rangle \gtrsim 1~\text{GeV}$), and high statistics. This can be achieved via beam luminosity, or target-mass considerations.  Fixed target and beam dump experiments---where neutrino energies can be in excess of $100~\text{GeV}$, and charged current event counts can exceed $10^6$---are an ideal setting to study neutrino trident production.  The Search for Hidden Particles (SHiP) experiment and the Deep Underground Neutrino Experiment (DUNE) both fall into these categories, and, as we show in this paper, represent the newest frontier in the study of trident production. 

SHiP's program of study, as it relates to neutrino physics, is largely focused on tau neutrino, and anti-tau neutrino events, and is therefore optimized to observe tau leptons \cite{SHiPCollaboration2015}. This represents a qualitatively new opportunity in the study of trident production, because the high mass of the tau leptons results in a threshold effect, wherein coherent production of a single tau lepton is not possible unless the following inequality holds $E_\nu>(1/2)m_\tau^2 R_{A}$; the bound for tau lepton pair production is given by $E_\nu>2 m_\tau^2 R_{A}$. As a result we also investigate the incoherent contribution to the cross section using both a diffractive and deep-inelastic approach. The experiment will use beams with $\langle E_\nu\rangle \approx 30~\text{GeV}-60~\text{GeV}$, and expects a lifetime collection of charged current events on the order of $N_{\text{CC}}\approx 2.7\cdot 10^{6}$ \cite{SHiPCollaboration2015}. It is therefore reasonable to assume that mixed flavour trident production, possibly including tau leptons, should be observable at the SHiP experiment. 

Although the focus of its program of study is neutrino oscillations, the Deep Underground Neutrino Experiment (DUNE) will use sufficiently high luminosities, and neutrino energies to induce trident production. DUNE consists of a near detector on site at FERMILAB \cite{DUNECollaboration2015} and a far detector at
Sanford Lab, both composed of liquid argon. This technology allows for the observation of both electrons, and muons. The far detector is exposed to a flux of neutrinos after a $1300~\text{km}$ transit through earth. The near detector will be used to account for systematic uncertainties in the neutrino beam and to record the initial neutrino flux. It is designed to obtain ten times the statistics of the far detector \cite{DUNECollaboration2015}. The expected charged current event count in the far detector over the lifetime of the experiment is on the order of $1 \cdot 10^5$, and so it is reasonable to expect an observable signal of trident events for some of the processes; especially given the enhanced statistics of the planned near detector.

Trident production has proven itself a useful tool for constraining BSM physics by virtue of its sensitivity to modifications of $C_A$ and $C_V$. Additionally it represents an experimental signal that would provide an obvious background to searches of lepton flavour violation in the case of multi-flavour charged-lepton tridents. If these new experiments (SHiP and DUNE) are to use trident production to probe BSM physics, then it is imperative to understand the relevant Standard Model backgrounds.

The rest of this article is organized as follows: In \cref{sec:Lepton-Matrix} we discuss the basic structure of the trident amplitude in the Standard Model. In \cref{sec:DIS-Vs-Coherent} we describe how to obtain the cross sections for three distinct kinematic regimes; each receiving a separate theoretical treatment. In \cref{sec:Prospects} we calculate expected rates, and cross sections for both DUNE and SHiP. We also present  differential distributions with respect to the invariant mass of the charged lepton pair. In \cref{sec:Discussion} we review the qualitative features of our results and outline possible applications of trident for both SHiP and DUNE. Finally in \cref{sec:Conclusion} we discuss future directions for trident production for the upcoming generation of accelerator based neutrino experiments.

\section{Trident Production in the Standard Model \label{sec:SM-Theory}}
\subsection{Leptonic Matrix Element \label{sec:Lepton-Matrix}}
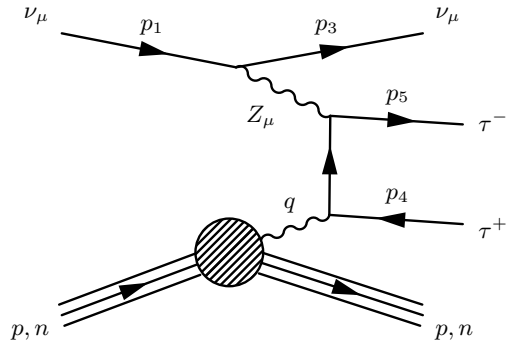
\begin{figure}[!ht]
	\centering
   \parskip 10pt   
    \qquad
	\begin{center}
		\begin{fmffile}{tryydent}
		\begin{fmfgraph*}(40,25)
		\fmfleft{i2,i1}
		\fmfright{o4,o3,o2,o1}
		\fmf{fermion,tension=7,label=$p_1$,label.side=left}{i1,v1}
		\fmf{fermion,tension=4}{i2,v4}
		\fmf{fermion,tension=3,label=$p_3$,label.side=left}{v1,o1}
		\fmf{photon,tension=7,label=$Z_\mu$,label.side=right}{v1,v2}
		\fmf{fermion,tension=5,label=$p_5$,label.side=left}{v2,o2}
		\fmf{fermion,tension=3,label.side=left}{v3,v2}
		\fmf{fermion,tension=5,label=$p_4$,label.side=right}{o3,v3}
		\fmfblob{.15w}{v4}
		\fmf{photon,tension=6.7,label=$q$,label.side=right}{v3,v4}
		\fmffreeze
		\fmfi{plain}{vpath (__i2,__v4) shifted (thick*(-0.5,2))}
		\fmfi{plain}{vpath (__i2,__v4) shifted (thick*(0.5,-2))}
		\fmf{fermion,tension=3}{v4,o4}
		\fmfi{plain}{vpath (__v4,__o4) shifted (thick*(-0.1,2))}
		\fmfi{plain}{vpath (__v4,__o4) shifted (thick*(-0.5,-2))}
		\fmflabel{$\nu_\mu$}{i1}
		\fmflabel{$p,n$}{i2}
		\fmflabel{$\nu_\mu$}{o1}
		\fmflabel{$\tau^-$}{o2}
		\fmflabel{$\tau^+$}{o3}
		\fmflabel{$p,n$}{o4}
	\end{fmfgraph*}
\end{fmffile}
	\end{center}
    \caption{An example of a process which takes place exclusively through the neutral current channel. The mismatch in flavour between the incident neutrino and outgoing leptons prohibits a charged current interaction.}
    \label{fig:tridentpicture}
\end{figure}

Our treatment of trident production varies over kinematic regimes, characterized by the four-momentum transfer to the nucleus $Q^2$. In every approach we treat the leptonic matrix element involving the EM current consistently. Our treatment of the nucleus' interaction with the EM field, however, varies, and so will be treated separately in each section. In the lower $Q^2$ regimes we relate the cross section to that of a neutrino-photon collision (photo-trident production), while for large $Q^2$ we employ the parton model. The amplitudes for photo-trident production and parton-trident production can be written
\begin{equation}
\begin{split}
\I\mathcal{M}_{\gamma\nu}&= \epsilon^\mu L_\mu ~~~~~~~~~~~~~\text{(EPA)} \\
\I\mathcal{M}_{h\nu}&= \frac{-\eta^{\mu\nu}}{q^2}  h_\nu  L_\mu~~~~~\text{(DIS)} 
\label{eq1:Lepton-Matrix}
\end{split}
\end{equation}
where $\epsilon^\mu$ is an on-shell polarization tensor, and $h_\nu$ is the hadronic matrix element in the parton model. The leptonic matrix element $L_\mu$ is calculated explicitly below.  We study both neutrino, and anti-neutrino induced trident production, and for the remainder of this section all reactions will contain an implicit hadronic initial and final state. We use Latin flavour indices  $i,j,k\in\{e,\mu,\tau\}$  and consider reactions of the form 
\[\{\nu_i \rightarrow\nu_{i~\text{or}~k} + \ell^-_{j} +\ell^+_{k}~,~~ \overline{\nu}_i \rightarrow \overline{\nu}_{i~\text{or}~j} +\ell^-_{j} + \ell^+_k\}
\]   
with the constraint that generational lepton number is conserved. Both mono-flavour, and multi-flavour charged lepton pairs (i.e. $\mu^+ \mu^-$ and $\mu^+\tau^-$) are included in our analysis. Assigning the labels $\{1,2,3,4,5\} \rightarrow\{\nu,\gamma,\nu',\ell^+,\ell^-\}$  with $\nu'$ the outgoing neutrino (see \cref{fig:tridentpicture}) and generalizing the analysis of \cite{ItayPrivate, Altmannshofer2014} to  multi-flavour lepton pairs we find 
 \begin{equation}
	\begin{split}
	L^\mu_{ijk}=-&\frac{\I e G_F}{\sqrt{2}}\left\{\overline{u}_3\gamma^\alpha (1-\gamma^5) u_1~~,~~\overline{v}_1\gamma^\alpha (1-\gamma^5) v_3\right\}\times\\
	&\overline{u}_5\bigg[\gamma_\alpha\left(V_{ijk}-A_{ijk}\gamma^5\right)\frac{1}{\slashed{q}-\slashed{p}_4-m_4}\gamma^\mu \\
	+&\gamma^\mu\frac{1}{\slashed{p}_5-\slashed{q}-m_5}\gamma_\alpha\left(V_{ijk}-A_{ijk}\gamma^5\right)\bigg]v(p_4),
	\end{split}
	\label{eq2:Lepton-Matrix}
\end{equation}
%
%
%
%
%
%
%
where the first line contains the appropriate spinor wavefunctions for an incident neutrino and anti-neutrino beam respectively. $V_{ijk}$ and $A_{ijk}$ are the flavour dependent vector and axial coupling strengths, which are typically denoted $C_V$ and $C_A$ respectively. We use non-standard notation to stress that these couplings carry flavour indices because some processes are mediated exclusively by $W$ bosons, others exclusively by $Z$ bosons, and some a mixture of the two. As we see from \cref{fig:tridentpicture}, these mediators modify the coupling to the vector and axial currents, as can be verified by use of Fierz identities. As noted in \cite{Belusevic1988} the interference between the neutral and charged current channels in the Standard Model results in a $40\%$ reduction in the cross section compared to the $V-A$ theory prediction. Thus by considering different combinations of leptons in the final state the cross section can be enhanced, or suppressed, significantly. The constants $A_{ijk}$ and $V_{ijk}$ are  presented in \cref{tab:neutrinoME} for $\nu_\mu\rightarrow \nu_\mu\tau^+\tau^-$ and for all trident processes with lifetime event counts greater than $0.01$ at either SHiP or DUNE. 

\begin{table}[t]
\centering
\begin{tabular}{ cccccc }
 \toprule
 $\nu$ Process				& $\overline{\nu}$ Process		 			& $V_{ijk}$ 			 & ~$A_{ijk}~$ 		& Mediator\\
\midrule
 $\nu_e\rightarrow \nu_e e^+e^-$ 	& $\overline{\nu}_e\rightarrow \overline{\nu}_e e^+e^-$		& $\frac{1}{2}+2\sin^2\theta_w$  & ~~$\frac{1}{2}$	& W,Z \\
 $\nu_\mu\rightarrow \nu_\mu \mu^+\mu^-$& $\overline{\nu}_\mu\rightarrow \overline{\nu}_\mu \mu^+\mu^-$	& $\frac{1}{2}+2\sin^2\theta_w$  & ~~$\frac{1}{2}$  	& W,Z \\
 $\nu_e\rightarrow \nu_\mu \mu^+e^-$	& $\overline{\nu}_e\rightarrow \overline{\nu}_\mu e^+\mu^-$	& $1$				 & ~~$1$			& W \\
 $\nu_\mu\rightarrow \nu_e e^+\mu^-$	& $\overline{\nu}_\mu\rightarrow \overline{\nu}_e \mu^+e^-$	& $1$				 & ~~$1$			& W \\
 $\nu_e\rightarrow \nu_e \mu^+\mu^-$	& $\overline{\nu}_e\rightarrow \overline{\nu}_e \mu^+\mu^-$	& $-\frac{1}{2}+2\sin^2\theta_w$ & $-\frac{1}{2}$	& Z \\
 $\nu_\mu\rightarrow \nu_\mu e^+e^-$	& $\overline{\nu}_\mu\rightarrow \overline{\nu}_\mu e^+e^-$	& $-\frac{1}{2}+2\sin^2\theta_w$ & $-\frac{1}{2}$	& Z \\
  $\nu_\mu \ra \nu_\mu \tau^+ \tau^-$ & $\overline{\nu}_\mu \ra \overline{\nu}_\mu \tau^- \tau^+$ & $-\frac{1}{2}+2\sin^2\theta_w$ &$-\frac{1}{2}$ & Z \\
 $\nu_\mu \ra  \nu_\tau\mu^- \tau^+$ & $\overline{\nu}_\mu \ra  \overline{\nu}_\tau\mu^+ \tau^-$ & 1 & ~~1 & W\\
 $\nu_\tau \ra \nu_\mu \tau^- \mu^+ $ & $\overline{\nu}_\tau \ra \overline{\nu}_\mu \tau^+ \mu^- $ & 1 & ~~1 & W\\
 $\nu_\tau \ra \nu_\tau\mu^+ \mu^-$ & $\overline{\nu}_\tau \ra \overline{\nu}_\tau\mu^- \mu^+$ & $-\frac{1}{2}+2\sin^2\theta_w$ & $-\frac{1}{2}$ & Z\\
  $\nu_\tau \ra \nu_\tau e^+ e^-$ & $\overline{\nu}_\tau \ra \overline{\nu}_\tau e^- e ^+$ & $-\frac{1}{2}+2\sin^2\theta_w$ & $-\frac{1}{2}$ & Z \\
\bottomrule
\end{tabular}
\caption{Modified vector and axial coupling constants for different combinations of incident neutrino flavours and final states}
\label{tab:neutrinoME}
\end{table}

\subsection{Coherent, Diffractive and Deep Inelastic Regimes \label{sec:DIS-Vs-Coherent}}
We will begin by reviewing conventional scattering of neutrinos off of nuclei to emphasize the qualitative differences in trident production. Neutrino-nucleus scattering is dominated by charged current events, which can be loosely partitioned into three classes for $E_\nu \gsim 100~\text{MeV}$: quasi-elastic scattering,  hadronic resonance production, and deep inelastic scattering \cite{Formaggio2012}. It is only at low centre of mass energies $E\lsim 50~\text{MeV}$ that coherent scattering via the neutral current is possible such that the reaction's cross section scales as $\sigma\thicksim(A-Z)^2E_\nu^2$ with $A-Z$ the number of neutrons. In this energy regime coherent scattering cross sections can be as much as three orders of magnitude larger than that predicted by a na{\"i}ve sum of the nucleon cross sections \cite{Drukier1984}. 

This limited kinematic window stands in sharp contrast to trident production where coherent contributions are possible at all energies, because the reaction is not $2\rightarrow 2$ and the phase space is therefore less kinematically constrained. This scattering is mediated electromagnetically, and, in addition to the coherent $Z^2$ amplification, the photon's propagator introduces an infrared divergence that further enhances the amplitude. As is the case for coherent neutrino scattering  this regime is characterized by small momentum transfers ($Q^2\sim R^{-2}_A$) wherein the phases of the various amplitudes are nearly commensurate, and the amplitudes interfere constructively. Kinematic considerations constrain the momentum transfer via $Q>s/(2E_\nu)$, with $s$ the invariant mass of the neutrino-photon pair \cite{Belusevic1988}. 
When combined with the lepton pair's mass threshold, this regulates the infrared divergence mentioned above. The three regimes typically considered in charged-current scattering for high energy neutrinos (mentioned in the first paragraph) also exist for trident production. Quasi-elastic-like diffractive scattering can contribute significantly to trident production, especially when threshold effects related to lepton masses are important. We expect the deep inelastic contribution to be suppressed, but for many of the neutrino energies at SHiP it is the only kinematically allowed production mechanism for tau leptons, and so we also include this regime in our analysis.


\subsubsection{Coherent Regime \label{sec:Coherent-Scattering}}

The coherent contribution to neutrino trident production can be accurately calculated using the equivalent photon approximation (EPA) \cite{Altmannshofer2014,Løvseth1971,Belusevic1988,Vysotsky2002}. In the EPA the cross section for the full scattering process is decomposed into two pieces. First the cross section corresponding to the scattering of a neutrino and photon creating a lepton trident, denoted by $\sigma_{\gamma\nu}$, is calculated. Next, this cross section is weighted against a universal probability distribution $P(s,Q^2)$ \cite{Altmannshofer2014} that measures the likelihood of the nucleus producing a virtual photon with virtual-mass $Q^2$, and neutrino-photon centre of mass energy $s$. The full cross section is given by 
\begin{equation}
\begin{split}
\sigma_{\nu A}&= \int \mathrm{d}s~  \sigma_{\gamma \nu}(s) \int \mathrm{d}Q^2 P(s,Q^2)\\
  &=\frac{Z^2\alpha}{\pi}
  \int_{m_{jk}^2}^{s_{\text{max}}}\frac{ \mathrm{d}s }{s}\sigma_{\gamma\nu}(s)
  \int_{(s/2E_\nu)^2}^\infty \frac{\mathrm{d}Q^2}{Q^2}F^2(Q^2)
  \label{eq2:Coherent-Scattering}
\end{split}
\end{equation}
with $m_{jk}=m_j+m_k$ the sum of the lepton pair's masses. 
A fairly good, albeit crude, approximation is to treat the form-factor for the nucleus $F(Q^2)$ as a Heaviside function $\Theta(Q^2_\text{max}-Q^2)$ where the scale  $Q_\text{max}=\Lambda_\text{QCD}/A^{1/3}$ corresponds to characteristic momentum transfer at which one would expect the dissolution of the nucleus \cite{Belusevic1988}. This sets a maximum centre-of-mass energy for the photon-neutrino interaction $s_\text{max}=2E_\nu Q_\text{max}$. With these approximations, suppressing flavour indices and working in the leading log approximation, \cref{eq2:Coherent-Scattering} simplifies to \cite{Belusevic1988,Altmannshofer2014} 
\begin{equation}
	\sigma_{\nu A} \approx \frac{1}{2}(A^2+V^2)\frac{2 ~Z^2\alpha^2 ~G_F^2 }{9\pi^3} s_\text{max}
   \log\left(\frac{s_\text{max}}{4m^2} \right)
	\label{eq3:Coherent-Scattering}
\end{equation}
where $2m=m_{j}+m_{k}$. There are additional terms resulting from the interference between the vector and axial currents, but these are suppressed by two powers of the lepton mass, and are therefore small. A more realistic implementation is to use the Woods-Saxon form-factor, which is what we used in all of our calculations (this changes the answer by order 10\%, see \cref{sec:Three-Body} for details). 
We can write the coherent contribution to the neutrino-nucleus cross section as 
\begin{equation}
	\mathrm{d}\sigma_{\gamma\nu}=\frac{1}{2s}\frac{1}{2} \sum_{\text{pol} }\left|\epsilon_\mu L^\mu \right|^2 \mathrm{d}\Phi_3
    \label{eq4:Coherent-Scattering}
\end{equation}
where $\Phi_3$ is the three-body phase space of final states, the factor of $1/2$ averages over photon polarizations, and  $2s$ is the Lorentz invariant flux factor. For details on the treatment of the three-body phase space see \cref{sec:Three-Body}.  

\subsubsection{Diffractive Regime}
At intermediate $Q^2$ it is possible to interact with the individual protons of the nucleus, both without coherent interference of their individual amplitudes, and without probing their inner parton structure. Our treatment of this regime follows the approach outlined in \cite{Brown1972}, and is identical to the coherent regime with the following changes: 
\begin{equation}
\begin{split}
\sigma_{\nu A}&=Z\int\mathrm{d}s~  \sigma_{\gamma \nu}(s) \int \mathrm{d}Q^2 P(s,Q^2)\\
  &=Z\frac{\alpha}{\pi}
  \int_{m_{jk}^2}^{s_{\text{max}}}\frac{ \mathrm{d}s }{s}\sigma_{\gamma\nu}(s)
  \int_{Q^2_\text{min}}^{1~\text{GeV}^2} \frac{\mathrm{d}Q^2}{Q^2}F_\text{dip}^2(Q^2).
  \label{eq1:Diffractive-Scattering}
\end{split}
\end{equation}
The charge of the nucleus now appears as an overall multiplicative factor as opposed to appearing in $P(s,Q^2)$, we cut off our integral at $Q_\text{min}=\text{max}\left(s/2E_\nu,R_A^{-1}\right)$ to avoid double counting amplitudes included in the coherent calculation, and we use the standard dipole fit to the proton's electromagnetic form factor (see \cref{sec:Three-Body}). We introduce an explicit UV cut-off for the $Q^2$ integration to avoid double counting with the DIS amplitudes. This was not necessary for the coherent regime due to the exponential, as opposed to power law, decay of the Wood-Saxon form factor at high $Q^2$.

\subsubsection{Deep Inelastic Regime} 
Our treatment of the deep inelastic case is fairly standard, with a few exceptions that are highlighted in \cref{sec:DIS-Hadron}. 
We treat this regime by convoluting the parton cross sections with nucleon parton distribution functions (PDFs) $f(\xi,Q)$ \cite{Martin2009}, taking into account the $u,d,c,s$ quarks. The phase space integrals are sensitive to the lepton masses, and so although their effects on the matrix element are often sub-leading, we include their full dependence throughout our calculations. All of the quarks are treated as massless in our analysis.

We take care to include a cut on momentum transfers so as not to double count contributions already accounted for by the EPA. Additionally we place a cut on the momentum fraction $\xi$ to ensure the parton carries enough four-momentum to both be able to produce the appropriate pair of charged leptons and to satisfy the double-counting-cut on momentum transfer. The resulting cross sections for the various nucleons are then summed to obtain the scattering cross section with the nucleus. We can write $\sigma_{\nu A}$ as a weighted sum of the cross sections with the constituent nucleons 
\begin{equation}
	\sigma_{\nu A}=Z\sigma_{\nu p}+(A-Z)\sigma_{\nu n}.
\end{equation}
These can in turn be written in terms of the parton-level cross sections $\sigma_{h\nu}$ via
\begin{equation}
 \sigma_{\nu H}=\sum_h\int_{\xi_\text{min}}^1 \mathrm{d}\xi \int^{Q_\text{max}}_{Q_\text{min}} \mathrm{d}Q~ \frac{\mathrm{d}\sigma_{h\nu}}{\mathrm{d}Q}(\xi,Q) ~f^{(H)}_h(\xi,Q) 
\end{equation}
where $f^{(H)}_h(\xi,Q)$ is the PDF for parton $h$ in the nucleon $H\in\{n,p\}$. More details can be found in \cref{sec:Four-Body}.

\section{Prospects at Future Experiments \label{sec:Prospects}}
In the following, we calculate trident rates at SHiP, and at the DUNE far and near detectors. We calculate the rates for momentum transfers $Q<0.217/(A)^\frac{1}{3}\GeV\approx R_A^{-1}$ regime using the coherent EPA method. For intermediate momentum transfers $0.217/(A)^\frac{1}{3}\GeV\lsim Q\lsim M_p$ transfers, we use the diffractive EPA treatment. Finally for  $Q\gsim1~\text{GeV}\approx  M_p $ we employ the deep inelastic formalism. We use PDFs from the MSTW collaboration (2008 NNLO best fit) \cite{Martin2009}. To calculate the rates, we estimate the number of SM neutrino trident events for each flavour of incident neutrino $\nu_i$ producing a lepton pair composed of $j^-$ and $k^+$ with $i,j,k\in\{e,\mu,\tau\}$. We estimate the luminosity in terms of charged current events  $N_{CC}^i$  using 
\be
	N^{ijk}_\text{Trident}=\sum_E \frac{N^i_{CC}(E)}{\sigma_{CC}(E,A)}\sigma^{ijk}_{\nu A}(E,Z,A)\times \epsilon^j_-\times \epsilon^k_+,
    \label{eq1:Prospects}
\ee
where $\sigma_{CC}$ is the neutrino charged current cross sections \cite{Agashe2014} and $i,j,k$ are flavours denoting the incident neutrino, outgoing $\ell^-$ and outgoing $\ell^+$ respectively. Additionally $\epsilon_+$ and $\epsilon_-$ are the identification efficiencies for $\ell^+$ and $\ell^-$ respectively. We do an analogous procedure for anti-neutrinos. 

There will be a background contribution to trident from resonant production of charged pions and charm production from $D$ mesons, whose leptonic modes are both dominated by muon flavoured final states. In the different flavour opposite sign di-lepton final states, backgrounds can arise from $\bar{\nu}_\mu$ CC scattering in combination with an elastic NC event releasing an electron, and also by muon final states in which one of the muons fake an electron. As coherent-scattering is quasi-elastic, the backgrounds for the dominant contribution to the cross section (see \cref{sec:DIS-Vs-Coherent}) can be greatly reduced by imposing hadronic vetoes in the analysis. Further background suppression can be achieved by selecting oppositely charged leptons that fall within the vertex resolution of the detectors and selecting events with low $M_{\ell^+\ell^-}$ invariant masses. We leave the background estimates to the collaborations' detailed and sophisticated simulations. Our signal results are shown in \cref{tab:ratesship,tab:ratesdunefar,tab:ratesdunenear}.
 \begin{table}[!th]
\centering
$\begin{array}{ ccccccc }
\multicolumn{3}{c}{\text{Neutrino Beam}}    &
\multicolumn{1}{c}{~} &
\multicolumn{3}{c}{\text{Anti-Neutrino Beam}}    \\ 
\cmidrule(lr){1-3}
\cmidrule(lr){5-7}
\text{Process}	  & \text{Coh}&\text{Diff}	    &~~&  \text{Process}	 			&\text{Coh}	& \text{Diff} \\
\midrule
\nu_{\mu}\ra\nu_ee^+\mu^-&85.46&24.6&\text{}&\bar{\nu}_{\mu}\ra\bar{\nu}_ee^-\mu^+&29.96&9.61\\
\nu_{\mu}\ra\nu_{\mu}e^+e^-&28.28&5.32&\text{}&\bar{\nu}_{\mu}\ra\bar{\nu}_{\mu}e^+e^-&22.48&3.58\\
\nu_e\ra\nu_ee^+e^-&21.69&2.95&\text{}&\bar{\nu}_e\ra\bar{\nu}_ee^+e^-&15.65&2.45\\
\nu_e\ra\nu_{\mu}\mu^+e^-&9.1&2.31&\text{}&\bar{\nu}_e\ra\bar{\nu}_{\mu}\mu^-e^+&14.31&3.16\\
\nu_{\mu}\ra\nu_{\mu}\mu^+\mu^-&4.79&3.01&\text{}&\bar{\nu}_{\mu}\ra\bar{\nu}_{\mu}\mu^+\mu^-&3.76&2.38\\
\nu_e\ra\nu_e\mu^+\mu^-&0.42&0.16&\text{}&\bar{\nu}_e\ra\bar{\nu}_e\mu^+\mu^-&0.3&0.12\\
\nu_{\tau}\ra\nu_{\tau}e^+e^-&0.13&0.03&\text{}&\bar{\nu}_{\tau}\ra\bar{\nu}_{\tau}e^+e^-&0.13&0.02\\
\nu_{\tau}\ra\nu_{\tau}\mu^+\mu^-&0.01&0.&\text{}&\bar{\nu}_{\tau}\ra\bar{\nu}_{\tau}\mu^+\mu^-&0.01&0.\\
\nu_{\tau}\ra\tau^-\mu^+\nu_{\mu}&0.&0.01&\text{}&\bar{\nu}_{\tau}\ra\tau^+\mu^-\bar{\nu}_{\mu}&0.&0.\\
\nu_{\mu}\ra\mu^-\tau^+\nu_{\tau}&0.&0.23&\text{}&\bar{\nu}_{\mu}\ra\mu^+\tau^-\bar{\nu}_{\tau}&0.&0.39\\
 \bottomrule
 \midrule
\text{Total}&149.88&38.62&\text{}&\text{}&86.6&21.71\\
 \bottomrule
\end{array}$
\caption{Number of expected trident events for coherent (Coh) and diffractive (Diff) scattering, using the EPA, in the \textbf{SHiP} $\nu_\tau$ detector, assuming $2\times 10^{20}$ POT on molybdenum. }
\label{tab:ratesship}
\end{table}

\subsection{Calibrations and Tests}
The details of our calculations can be found in the Appendices. We calibrated our EPA cross section calculations with previous theoretical and experimental work \cite{Altmannshofer2014,Geiregat1990,Geiregat1993}, and reproduced the analytic results of \cite{Altmannshofer2014}.

Our DIS work was calibrated with MadGraph5 \cite{Alwall2014} for trident induced muon pair production. MadGraph5 treats light leptons as massless, and due to infrared singularities in the propagators this necessitates a careful treatment; it also introduces questions of reliability. We imposed the following cuts to replicate the effects of finite muon masses: $p_{T}>m_\mu$ for the muons, $p_{T}>1.5~\text{GeV}$ for the jets, and $\Delta R=\sqrt{\Delta\eta^2+\Delta\phi^2}>0.4$ for the lepton pairs. With these cuts we found our calculations to agree with MadGraph5 to within a factor of $0.5-2.5$ for $E_\nu=\{20~\text{GeV},200~\text{GeV},1000~\text{GeV}\}$. We believe our calculation to be more reliable than MadGraph5 in the low $Q^2$ regions of phase space which dominate the cross sections due to infrared divergences, which we treat carefully.

\subsection{Rates for SHiP}

SHiP will be a lead based neutrino detector \cite{Bonivento2013,SHiPCollaboration2015}. It will utilize an emulsion cloud chamber for its electron detection and a muon magnetic spectrometer for muons. It is estimated to have a 90\% $e$ and $\mu$ identification efficiency,
 and a micron vertex resolution. Under nominal operating conditions, after 5 years of operation it will have collected data from $2\times 10^{20}$ POT using a 400$\GeV$ SPS proton beam. We quote all the rates assuming this normalization. 

\begin{figure}[h]
	\centering
	\includegraphics[width=0.48\textwidth]{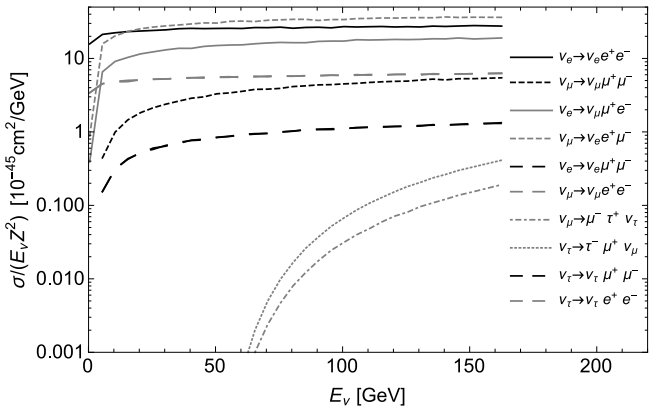}
	\caption{$\sigma/E_\nu$ trident cross sections normalized by $Z^2$ for various SM flavours as a function of the incoming neutrino energy on a lead target (SHiP).}
	\label{fig:smsigmavseshipEPA}
\end{figure}

The energy spectrum at SHiP is very broad, and reaches sufficiently high energies such that trident production of tau leptons becomes kinematically allowed in the coherent, diffractive, and deep inelastic regimes. The latter is allowed at almost all incident neutrino energies available at SHiP with the only requirement being the centre of mass energy exceed the lepton pair's mass-gap. Despite being kinematically allowed, we find the large momentum transfer in the deep  inelastic regime renders the contribution to the cross section negligible. The diffractive and coherent regimes rely on the high energy tail of the quoted beam distribution \cite{SHiPCollaboration2015}. For electrons and muons, coherent, and diffractive production are not only possible but extremely viable, while for tau leptons we find only diffractive production to be viable, but only marginally so. In \cref{fig:smsigmavseshipEPA} and \cref{fig:smsigmavseshipDIS}, we show the cross section per nucleon as a function of the incoming neutrino energy for a variety of processes. The coherent cross sections computed via the EPA are normalized by $Z^2$ while the deep inelastic contribution is normalized by $A$. There are small differences in these plots for various materials, as the EPA Woods-Saxon form factor and the relative number of protons to neutrons in DIS both introduce a sub-leading dependence on the ratio of protons to neutrons that is not removed by the per nucleon normalization.

In \cref{tab:ratesship} we show the expected number of events in the various production modes for both low-$Q^2$ events calculated within the coherent EPA and intermediate-$Q^2$ events calculated using the diffractive EPA. DIS rates are not included, because the cumulative lifetime event-count for all production modes in the deep inelastic regime is  $N^{(\text{tot})}_\text{DIS}\approx 0.1$. 

The basic features of our analysis can be understood by looking at \cref{tab:neutrinoME,eq3:Coherent-Scattering} and remembering that 
the neutrino beam is dominated by $\nu_\mu$ and $\overline{\nu}_\mu$. This is discussed in greater detail in \cref{sec:Discussion}.

\subsection{Rates for DUNE}
 \begin{table}[h]
\centering
$\begin{array}{ ccccccc }
\multicolumn{3}{c}{\text{Neutrino Beam}}    &
\multicolumn{1}{c}{~} &
\multicolumn{3}{c}{\text{Anti-Neutrino Beam}}    \\ 
\cmidrule(lr){1-3}
\cmidrule(lr){5-7}
\text{Process}	  & \text{Coh}&\text{Diff}	    &~~ &  \text{Process}	 			&\text{Coh}	& \text{Diff} \\
\midrule
\nu_{\mu}\ra\nu_ee^+\mu^-&73.98&53.15&\text{}&\bar{\nu}_{\mu}\ra\bar{\nu}_ee^-\mu^+&25.23&18.7\\
\nu_{\mu}\ra\nu_{\mu}e^+e^-&23.03&9.64&\text{}&\bar{\nu}_{\mu}\ra\bar{\nu}_{\mu}e^+e^-&16.45&6.79\\
\nu_{\mu}\ra\nu_{\mu}\mu^+\mu^-&2.03&5.28&\text{}&\bar{\nu}_{\mu}\ra\bar{\nu}_{\mu}\mu^+\mu^-&2.16&4.3\\
\nu_e\ra\nu_ee^+e^-&0.7&0.29&\text{}&\bar{\nu}_e\ra\bar{\nu}_ee^+e^-&0.54&0.22\\
\nu_e\ra\nu_{\mu}\mu^+e^-&0.21&0.17&\text{}&\bar{\nu}_e\ra\bar{\nu}_{\mu}\mu^-e^+&0.4&0.27\\
\nu_e\ra\nu_e\mu^+\mu^-&0.01&0.01&\text{}&\bar{\nu}_e\ra\bar{\nu}_e\mu^+\mu^-&0.&0.01\\

\bottomrule
 \midrule
\text{Total}&99.96&68.54&\text{}&\text{}&44.78&30.29\\
\bottomrule
\end{array}$
\caption{Number of expected trident events for coherent (Coh) and diffractive (Diff) scattering, using the EPA, in the lifetime of the \textbf{DUNE near} detector assuming $\sim 3\times 10^{22}$ POT (equivalently, an $850~\text{kt-MW-yr}$ exposure at the far detector).}
\label{tab:ratesdunenear}
\end{table}

DUNE \cite{DUNECollaboration2015} is composed of a near detector that primarily sees a flux of muon neutrinos and a far detector used to study the appearance of electron neutrinos as a result of oscillations from the muon neutrino beam. That said, there will be a mixture of both neutrino flavours at each site relevant for trident. Both near and far detectors are based on argon time projection chambers, which allow for the differentiation of electrons and photons. We take the electron and muon identification efficiencies to be 90\%.

In \cref{tab:ratesdunenear} and \cref{tab:ratesdunefar}, we show the expected number of events for the near and far detectors respectively. The rates in both tables are calculated assuming an 850kt-MW-yr exposure in the far detector. This number corresponds to the amount of data collected in the lifetime of DUNE given their optimized design. To convert this measure to protons on target, note that the far detector weighs 40kt, and a beam power of 1.07MW with 80GeV protons corresponds to $1.47\times 10^{21}$ POT/yr \cite{DUNECollaboration2015}. This gives roughly $3\times 10^{22}$ POT. The full details of the luminosity calculations are given in \cref{sec:luminosity}. As we did for SHiP, we consider both low-$Q^2$ events calculated within the coherent EPA and intermediate-$Q^2$ events calculated using the diffractive EPA. DIS rates are not included as they are negligible. In \cref{fig:smsigmavsedune}, we show the cross section per nucleon as a function of the incoming neutrino energy for each process listed in \cref{tab:neutrinoME}, for coherent EPA. Comparing to \cref{fig:smsigmavseshipEPA} there are small differences, which are due to the Woods-Saxon form factor's implicit dependence on $A$ (see \cref{eq8:Three-Body} for details).

\begin{table}[!h]
\centering
$\begin{array}{ ccccccc }
\multicolumn{3}{c}{\text{Neutrino Beam}}    &
\multicolumn{1}{c}{~} &
\multicolumn{3}{c}{\text{Anti-Neutrino Beam}}    \\ 
\cmidrule(lr){1-3}
\cmidrule(lr){5-7}
\text{Process}	  & \text{Coh}&\text{Diff}	    &~~ &  \text{Process}	 			&\text{Coh}	& \text{Diff} \\
\midrule
\nu_{\mu}\ra\nu_ee^+\mu^-&2.12&1.52&\text{}&\bar{\nu}_{\mu}\ra\bar{\nu}_ee^-\mu^+&0.05&0.03\\
\nu_{\mu}\ra\nu_{\mu}e^+e^-&0.66&0.28&\text{}&\bar{\nu}_{\mu}\ra\bar{\nu}_{\mu}e^+e^-&0.03&0.01\\
\nu_e\ra\nu_ee^+e^-&0.11&0.05&\text{}&\bar{\nu}_e\ra\bar{\nu}_ee^+e^-&0.05&0.02\\
\nu_{\mu}\ra\nu_{\mu}\mu^+\mu^-&0.06&0.15&\text{}&\bar{\nu}_{\mu}\ra\bar{\nu}_{\mu}\mu^+\mu^-&0.&0.01\\
\nu_e\ra\nu_{\mu}\mu^+e^-&0.03&0.03&\text{}&\bar{\nu}_e\ra\bar{\nu}_{\mu}\mu^-e^+&0.03&0.02\\
 \bottomrule
 \midrule
\text{Total}&2.98&2.03&\text{}&\text{}&0.16&0.09\\
 \bottomrule
\end{array}$
\caption{Number of expected trident events for coherent (Coh) and diffractive (Diff) scattering, using the EPA, in the lifetime of the \textbf{DUNE far} detector assuming $\sim 3\times 10^{22}$ POT (equivalently, an $850~\text{kt-MW-yr}$ exposure at the far detector).
\label{tab:ratesdunefar}}
\end{table}

\section{Discussion and Analysis \label{sec:Discussion}}

The general features of \cref{sec:Prospects} can be understood qualitatively by considering \cref{eq3:Coherent-Scattering} and \cref{tab:neutrinoME}. First we note that every cross section is proportional to the combination $\left|C_V\right|^2+\left|C_A\right|^2$ appearing in \cref{eq3:Coherent-Scattering}. In the SM this is maximal in the case of W mediated interactions, intermediate for W$+$Z mediated interactions, and minimal for Z mediated interactions. The W exclusive channel corresponds to scattering events where the incoming and outgoing neutrino belong to different lepton generations, and thus these channels will be more probable. Another dominant feature controlling the relative size of cross sections is related to the masses of the outgoing leptons. This dictates the size of the logarithmic enhancement coming from the low $Q^2$ phase space. This is a feature of the IR divergence arising from the photon propagator, which is regulated by the finite masses of the charged leptons. Finally the rates quoted in \cref{tab:ratesship,tab:ratesdunefar,tab:ratesdunenear} are further influenced by beam luminosity, and so tend to favour incident muon configurations, except at the DUNE far detector, where they favour incident electron neutrinos. 

\begin{figure}
	\centering
	\includegraphics[width=0.48\textwidth]{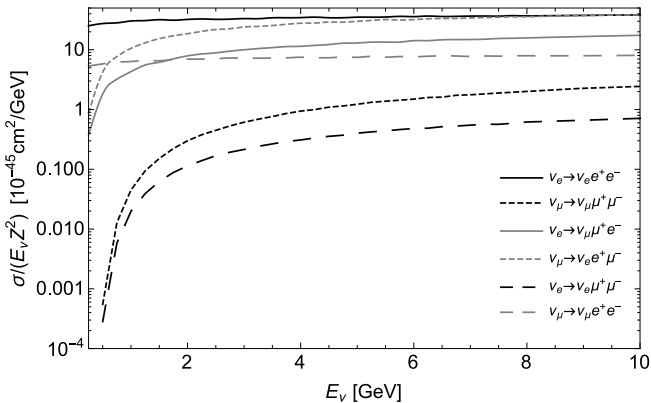}
	\caption{$\sigma/E_\nu$ trident cross sections normalized by $Z^2$ for various SM flavours as a function of the incoming neutrino energy on an argon target (DUNE).}
	\label{fig:smsigmavsedune}
\end{figure}

These qualitative features suggest that  $\nu_\mu\rightarrow \nu_e\mu^-e^+$ would serve as the dominant production mode at both the DUNE near detector and SHiP. Examining \cref{tab:ratesdunenear,tab:ratesship}, this is indeed the case. It is a CC-exclusive process (high axial-vector couplings), it benefits from the large flux of muon neutrinos, and from the logarithmic enhancement afforded by the low electron mass. This final statement is most important at DUNE due, to its lower $\langle E_\nu \rangle$, which makes it sensitive to muon-mass threshold effects. For diffractive processes the sensitivity of the cross section to the charged lepton masses is weakened due to the lower bound $Q_\text{min}$ in \cref{eq1:Diffractive-Scattering}. This accounts for the difference in ordering of rates between the coherent and diffractive contributions to the cross section found in \cref{tab:ratesdunenear,tab:ratesdunefar,tab:ratesship}. At DUNE this results in an enhancement of the cross section by a factor of $35$ when compared to the production mode $\nu_\mu\rightarrow\nu_\mu\mu^+\mu^-$, which was observed at CHARM-II, CCFR, and NuTeV \cite{NuTeVCollaboration1998,Geiregat1993,Geiregat1990}. No dedicated search was carried out for electron production in trident modes at these experiments. The detector technology typically consisted of interwoven layers of heavy element materials to induce neutrino interactions, followed by calorimeters to measure the final lepton states. Electrons create showers and scatter much more in these layers, as opposed to muons which tend to follow a straight trajectory until the muon spectrometer. It was thus much more difficult to impose vertex requirements on electrons, which is an integral part of the trident analysis. Neutrino detector technology has greatly evolved since then, and it is now feasible to consider mixed flavour trident channels.

The lifetime expected event count for $\mu^+\tau^-$ and $\mu^-\tau^+$ production are both approximately unity. Given the uncertain run-time and technical specifications of SHiP it is possible that tridents containing tau leptons will occur,  however the rates are sufficiently low that it is not clear at what level of statistical significance these can be observed, especially after applying necessary cuts. Our analysis suggest that these events are most likely to occur for intermediate momentum transfers (i.e. in the diffractive regime). Our deep inelastic analysis revealed high-$Q^2$ trident production to be extremely suppressed for all flavours, including tau leptons. $\nu_\tau$ induced electon-muon pairs may be observable, however, due to the much higher flux of $\nu_\mu$'s this channel will be dominated by $\nu_\mu$ induced events with identical charged lepton final states, which will leave an indistinguishable signature in the detector.

In the case of the DUNE collaboration, the size of the near detector is currently being planned such that it can obtain approximately ten times the statistics of the far detector; allowing for a reduction in the systematic uncertainties of the neutrino beam. Our results show that even for near detector masses that minimally satisfy this requirement trident production should be detectable. Given the large beam intensity at the near detector, every additional unit of detector mass represents a fantastic return on investment from the perspective of rare neutrino processes such as trident production. Pushing from hundreds to thousands of events would lower statistical error to the level of a few percent, and could potentially allow for trident production to act as a complimentary beam characterization tool. This is alluring because trident production is only sensitive to the target nucleus' electric form factor, in contrast to CC events where uncertainties in the axial form factor still introduce significant systematic effects. 

While interesting in its own right as a test of the Standard Model, neutrino trident production can also act as a significant background in the search for new physics. This is because of its qualitative similarities to processes involving lepton flavour violation, which is a signature of many BSM models. Our estimated rates also suggest that both SHiP and the DUNE near detector can be used to constrain BSM physics; comparison with the number of events identified by the CCFR, and CHARM-II collaboration in the di-muon channel alone demonstrates that both SHiP and DUNE are competitive with these previous experiments. With access to flavour dependent final states, however, we believe these experiments can do much better. For example the $Z'$ coupling to $L_\mu-L_\tau$ considered in \cite{Altmannshofer2014} influences both $\nu_\mu\rightarrow \nu_\mu \mu^+ \mu^-$ and $\nu_\mu\rightarrow\nu_\mu e^+ e^-$. Due to the minimal size of $\left|C_V\right|^2 +\left|C_A\right|^2$ for $e^+e^-$ production (due to Z-exclusive mediation) this process will experience an even greater relative sensitivity to new physics, albeit in a first-generation lepton channel.

Although the qualitative features discussed earlier are sufficient to understand the most prominent aspects of our analysis, a closer examination of \cref{fig:smsigmavsedune,fig:smsigmavseshipEPA} reveals another feature, which is initially surprising. The rates for processes which seem to be related by an exchange of flavour indices have different cross sections. This effect is $\mathcal{O}(1)$ and independent of energy (see \cref{fig:smsigmavseshipEPA} $\nu_\mu\rightarrow\nu_\tau \tau^+\mu^-$ vs $\nu_\tau\rightarrow\nu_\mu\mu^+\tau^-$ for example). This would seem to suggest a violation of lepton universality, however a closer examination reveals that the chiral structure of the outgoing leptons is not equivalent, with the amplitudes for production into inequivalent configurations being proportional to the square of the heaviest lepton mass. Still this effect is surprising given that it is independent of energy, and na\"ively one would expect that at sufficiently high centre of mass energies the effect would be suppressed by $m_\ell^2/S$ with $S$ the Mandelstam variable for the neutrino-nucleus interaction. This is not the case for trident production because the cross section is dominated by the low-$Q^2$ region of phase space. To understand this we turn to the EPA, and more specifically \cref{eq2:Coherent-Scattering}. We see that the integral over $s$ has an IR cutoff of $m^2_{ij}=(m_{\ell^+}+m_{\ell^-})^2$, and so in this regime we find an $\mathcal{O}(m_{ij}^2/s)\thicksim\mathcal{O}(1)$ contribution to the cross section, which will be present even for arbitrarily high $E_\nu$. 

\begin{figure}
  \includegraphics[width=0.45\linewidth]{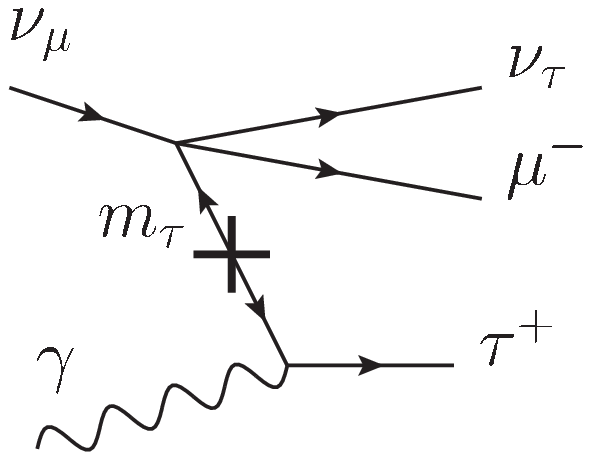}
  \includegraphics[width=0.45 \linewidth]{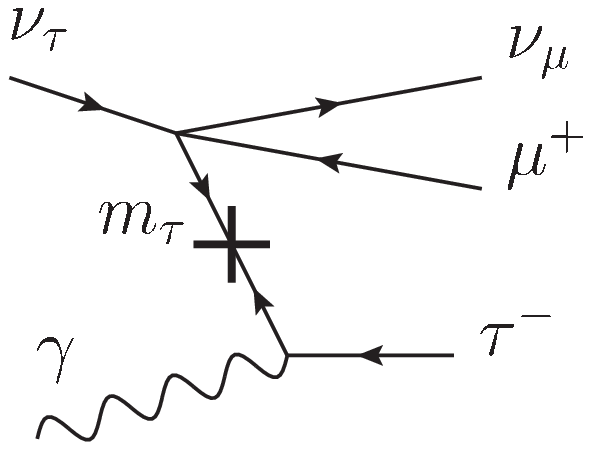}
  \caption{Inequivalent contributions to the processes $\nu_\mu\rightarrow \nu_\tau\tau^+\mu^-$ (left) and $\nu_\tau\rightarrow\nu_\mu\mu^+\tau^-$ (right) in the limit of $m_\mu\rightarrow0$. Note that 
the chiral structure of the weak interaction results in a triplet of left-handed leptons (LLL) for incident $\nu_\mu$ and a right-handed lepton pair with a left-handed neutrino (LRR) for incident
$\nu_\tau$. The fermions are two component spinors of definite chirality. Diagramatic conventions  are from \cite{Dreiner:2008tw} with arrows denoting chirality. \label{fig:chiral-diagram}}
\end{figure}

To understand why the chiral structure of the amplitude has a significant 
impact on the amplitude we must consider both the infrared divergence of the 
photon mediator, and the constraints imposed by conservation of angular 
momentum. Consider the centre of mass frame for the photon-neutrino pair.
To  saturate the lower bound of the integral over $s$ in 
\cref{eq2:Coherent-Scattering} we must produce the lepton-pair at rest, and have the 
neutrino red-shift to an arbitrarily small energy
$E'_\nu=\epsilon$; this also forces the lepton pair to carry equal an opposite
momentum. It is, however, difficult to understand the implications of chirality
in this frame, because in this frame the lepton pair is non-relativistic and 
we cannot freely interchange helicity and chirality. 

To solve this problem we can appeal to Lorentz invariance and perform the same 
analysis in a boosted frame in which the lepton-pair is highly relativistic. 
To do this boost in the direction of infinitesimal momentum for the lepton 
pair. This boost will further red-shift the outgoing neutrino, but it will not
change its direction. We would like to check if this configuration conserves
angular momentum, and the answer to this question is dependent on the initial 
polarization of the incident photon (the neutrino's polarization is fixed because of its definite chirality), which in turn determines the initial angular momentum. The two 
possibilities are $S=1/2$ and $S=3/2$.

As shown in \cref{fig:chiral-diagram} the outgoing states for the two 
configurations have different chirality (LLL vs LRR). In the case of 
$S=1/2$ this has no effect on the configuration discussions above, however in the
case of $S=3/2$, where the spin of the neutrino and photon are aligned, 
the LLL configuration is forbidden, while the LRR configuration is allowed. 
This is because in our boosted frame, where chirality is equivalent to helicity, in order to obtain $S=3/2$ for the LLL configuration all three particles would have to 
travel in the same direction, which would violate conservation of momentum. 
Thus only the LRR, and not the LLL, configuration satisfies all the necessary conservation laws in the low-$Q^2$ region of phase space that dominates 
\cref{eq2:Coherent-Scattering}.  

\section{Conclusions \& Outlook \label{sec:Conclusion} }
We have demonstrated that as of yet unobserved neutrino trident
processes are within reach with the planned DUNE and SHiP experimental
collaborations. The DUNE collaboration may be able to enhance
production modes, some of which we currently estimate to only yield
$1-10$ events in the experiment's lifetime
(e.g. $\nu_\mu\rightarrow\nu_\mu\mu^+\mu^-$), 
by increasing the mass of the relatively small
near detector. Even with the current proposed designs both
collaborations are maximally sensitive to the mode
$\nu_\mu\rightarrow\nu_e\mu^-e^+$ and $\overline{\nu}_\mu\rightarrow
\overline{\nu}_e \mu^+e^-$. We  believe that backgrounds for these
searches will be low, especially given the vertex resolution of both
experiments \cite{SHiPCollaboration2015,DUNECollaboration2015}.

In addition to our direct application to the DUNE and SHiP
collaboration we also present $\sigma(E_\nu)$ for the coherent
scattering regime, allowing for future analyses with more precise
luminosity estimates. We present a similar plot in
\cref{fig:smsigmavseshipDIS} in case high momentum-transfer trident is
of future interest. We have considered all possible combinations of
lepton flavour final states, and have presented only processes with
non-zero lifetime event counts. This work is complementary to that
found in \cite{Løvseth1971}, where differential distributions with
respect to the lepton pair's invariant mass are plotted in the
coherent regime. Additionally we have demonstrated a method for
treating neutrino trident production on the parton level, which
requires some slight modifications to the standard treatment. This
revealed high-$Q^2$ trident production is untenable as one would
na\"ivey expect. 

Neutrino trident production is a proven tool in the testing of the SM
and constraining BSM physics, and with improved detector designs it is
important to harness the full capabilities of next generation neutrino
experiments.
Our analysis suggests that both SHiP and DUNE will be able to observe
trident production. We believe with these experiments on the horizon
the future is bright for studying trident production and other rare
neutrino processes, and that the study of these processes should be
incorporated into the physics programs of both experiments.

\section*{Acknowledgements}
We are very grateful to Maxim Pospelov for suggesting mixed flavour trident production and its applicability to future intensity frontier experiments. Additionally we would like to thank him for his continued guidance throughout this research. We would also like to thank Itay Yavin for his help in the treatment of phase space. Finally we thank Chien-Yi Chen and Richard Hill for useful discussions. This research was supported in part by Perimeter Institute for Theoretical Physics. Research at Perimeter Institute is supported by the Government of Canada through the Department of Innovation, Science and Economic Development and by the Province of Ontario through the Ministry of Research and Innovation. This research was also supported by funds from the National Science and Engineering Research Council of Canada (NSERC), the Ontario Graduate Scholarship (OGS) program, and the Early Research Awards program of Ontario.

\begin{appendices}
\appendix
\section{3-Body Phase Space (EPA) \label{sec:Three-Body}}
For the purposes of the EPA, the phase space integrals are performed
over the 3-body phase space of the leptons. Ultimately this 3-body phase 
space is embedded in the full 4-body one, and so we will use the results of 
this section in the proceeding one.  We denote the centre of mass energy for the photon-neutrino collision by $s$, additionally we define the quantities 
$P=p_+ + p_-$ and $\ell=P^2$. We begin by decl omposing the 3-body phase space 
using the identity below.
\begin{equation}
  \mathrm{d}\Phi_3(p_+,p_-,k_2)=
  \frac{\mathrm{d}\ell}{2\pi}\Phi_2(k_2,P)\Phi_2(p_+,p_-).
  \label{eq1:Three-Body}
\end{equation}
Each two-body phase space can be expressed as 
\begin{equation}
  \mathrm{d}\Phi_2(q_1,q_2)=\overline{\beta}(q_1,q_2)
  \frac{\mathrm{d}\Omega}{32\pi^2}
  \label{eq2:Three-Body}
\end{equation}
with the definition 
\begin{equation}
  \overline{\beta}(q_1,q_2)=
   \sqrt{1-\frac{2\left(q_1^2+q_2^2\right)}{(q_1+q_2)^2}+ \
        \frac{\left(q_1^2-q_2^2\right)^2}{(q_1+q_2)^4}}
	\label{eq3:Three-Body}
\end{equation}
An important case is when $q_1^2=0$. In this scenario the
factor simplifies to 
$\overline{\beta}=1-\frac{q_2^2}{ (q_1 + q_2 )^2}$. 
In our decomposition above $\overline{\beta}(k_2,P)=1-\ell/s$. 
First we choose to evaluate $\mathrm{d}\Phi_2(P,k_2)$ in the centre of mass
frame of the reaction. This allows us to parameterize the phase-space as 
written in \cref{eq2:Three-Body}. We can perform the azimuthal integration by
appealing to symmetry, and we are left only with 
$\mathrm{d}\cos{\theta_{CM}}$. This can conveniently be expressed in terms 
of the Lorentz-invariant $t$ defined via 
\begin{equation}
  t=2q_\mu(k_1-k_2)^\mu=\frac{1}{2}(s +\ell + ( \ell - s) \cos\theta_{CM}).
  \label{eq4:Three-Body}
\end{equation}
This definition leads to the differential relationship 
$\mathrm{d}t=\tfrac{1}{2}(\ell-s)\mathrm{d}\cos\theta$. Thus we can 
simplify our 3-body phase space integral by applying the identity
$\overline{\beta}(k_2,P)d\cos{\theta}=-\tfrac{2}{s}\mathrm{d}t$. 
This leaves us with the second phase space integral. This is most easily 
evaluated in the frame where $P_\mu$ has vanishing three-momentum. In this 
frame there is no guarantee of azimuthal symmetry in the matrix element, 
and so we must integrate over both polar angles. We are left with the 
expression quoted in \cite{Altmannshofer2014} 
\begin{equation}
  \mathrm{d}\Phi_3(k_2,p_+,p_-)=\frac{1}{2}\frac{1}{(4\pi)^2}
  \frac{\mathrm{d}\ell}{2\pi}\overline{\beta}(p_+,p_-)
  \frac{\mathrm{d}t}{2s}\frac{\mathrm{d}\Omega}{4\pi}
	\label{eq5:Three-Body}
\end{equation}
where we denote the angular integral over the muon-pair, performed in the
frame where $P=(\sqrt{\ell},0,0,0)$ by $\mathrm{d}\Omega$. The limits of 
integration for $t$ are given by $\ell<t<s$. This gives the expression for 
the photon-neutrino cross section as 
%
%
\begin{equation}
  \sigma_{\gamma\nu}=\frac{1}{2s}\frac{1}{2}\frac{1}{(4\pi)^2}
  \int_{m_{jk}^2}^s \frac{\mathrm{d}\ell}{2\pi}\overline{\beta}_\pm(\ell)
  \int_\ell^s\frac{\mathrm{d}t}{2s}\int\frac{\mathrm{d}\Omega}{4\pi}
 \left|\overline{\mathcal{M} }\right|^{2}_{\gamma\nu}
	\label{eq6:Three-Body}
\end{equation}
where $m_{jk}=m_j+m_k$, and ${\displaystyle \left|\overline{\mathcal{M} }\right|^{2}=1/2\sum_\text{pol}\left|\mathcal{M}\right|^2}$.

To obtain the full cross section this must be weighted against the 
probability for creating a photon in the Coulomb field of a nucleus, given in 
\cite{Belusevic1988,Altmannshofer2014}. This leads to 
\begin{equation}
  \sigma_{N\nu}=\frac{Z^2\alpha}{\pi}
  \int_{m_{jk}^2}^S\frac{ \mathrm{d}s }{s}\sigma_{\gamma\nu}(s)
  \int_{(s/2E_\nu)^2}^\infty \frac{\mathrm{d}Q^2}{Q^2}F^2(Q^2)
  \label{eq7:Three-Body}
\end{equation}
where $\sqrt{S}$ denotes the neutrino-nucleus centre of mass energy. In practice, the form factor of the nucleus $F(Q^2)$ cuts this integral off near $s_{\text{max}}\approx 2 E_\nu \Lambda_\text{QCD}/A^{1/3}$. In our calculations for the coherent regime (\cref{sec:Coherent-Scattering}) we used the Woods-Saxon form factor
\begin{equation}
F_\text{WS}(Q^2)=\frac{1}{N}\mathcal{F}\left\{ \frac{V_0}{1+\exp\left(\frac{r-r_0A^{1/3}}{a}\right)}\right\}
\label{eq8:Three-Body}
\end{equation}
with $\mathcal{F}$ denoting the Fourier transform with respect to $r$, and $N$ is a normalization factor given by the volume integral over the nuclear charge distribution\cite{Jentschura:2009mb}. The various parameters are set as $r_0\approx1.126~\text{fm}$, $a\approx0.523~\text{fm}$, and  $V_0=(4 \pi A r_0^3/3)^{-1}$. Different choices of form factor modify the result on the $10\%$ level.

For the diffractive regime we used the electric dipole fit for the proton's Dirac form factor found in \cite{Brown1972,Perdrisat2006,Formaggio2012}. Due to the quasi-elastic nature of the scattering the Pauli form factor's contribution is suppressed. The explicit expression is given by
\begin{equation}
F_\text{dip}(Q^2)=\frac{G_\text{dip}(Q^2)+\tau ~\xi G_\text{dip}(Q^2)}{1+\tau}
\label{eq9:Three-Body}
\end{equation}
where $\tau=Q^2/4M^2$ with $M=(m_p+m_n)/2$ and $\xi=(\mu_p-\mu_n)/\mu_N\approx 4.7$ the difference in magnetic moments between the proton and the neutron measured in units of the nuclear magneton. The dipole fit is given by
\begin{equation}
G_\text{dip}(Q^2)=\left(1+\frac{Q^2}{0.71~\text{GeV}^2}\right)^{-2}.
\label{eq10:Three-Body}
\end{equation}
%
%
\section{4-Body Phase Space (DIS) \label{sec:Four-Body}}
\subsection{Parton-Neutrino Collision \label{sec:DIS-Parton}}
We now consider the decomposition of the 4-body phase space. This will 
involve a reduction to the previously analyzed 3-body case, however there 
will be some difference thereafter because of the loss of azimuthal 
symmetry in
$\Phi_2(P,k_2)$. 

\begin{figure}
\includegraphics[trim={0 16.5cm 0 5.5cm},clip,width=\linewidth]{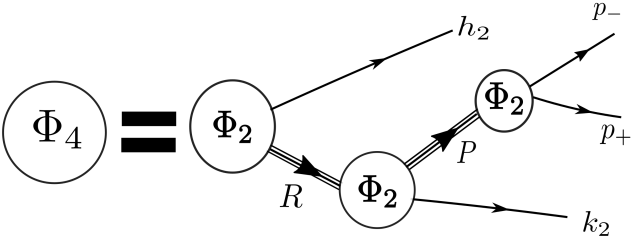}
\caption{Schematic depiction of the four-body phase space decomposition into three two-body phase spaces. Note the three-body phase space decomposition for the EPA is obtained by considering only the final two phase spaces in the diagram. \label{fig:schematic-phase-space}}
\end{figure}

We begin by emphasizing a change in notation. The centre of mass energy 
for the parton-neutrino collision is denoted $S$, we introduce the 
four-vector $R=k_2 + p_+ +p_-$ and its invariant mass $L=R^2$, and we 
maintain the previous definition of $P=p_-+p_+$.  We can now decompose the 
4-body phase space as shown schematically in 
\cref{fig:schematic-phase-space} and more precisely below:
\begin{equation}
  \begin{split}
  &~~~~\mathrm{d}\Phi_4(p_+,p_-,h_2,k_2) \\
  &=\frac{\mathrm{d}L}{2\pi}\mathrm{d}\Phi_2(R,h_2)\mathrm{d}\Phi_3(p_+,p_-,k_2)\\
  &=\frac{\mathrm{d}\ell}{2\pi}\frac{\mathrm{d}L}{2\pi} \mathrm{d}\Phi_2(R,h_2)
  \Phi_2(k_2,P) \Phi_2(p_+,p_-).
\end{split}
\label{eq1:Four-Body}
\end{equation}
The first two-body phase space $\Phi_2(h_2,R)$ inherits the azimuthal 
symmetry of the parton-neutrino collision, and in direct analogy with \cref{eq4:Three-Body} 
we introduce the variable $T$ defined via 
\begin{equation}
  T=2h_1^\mu(k_1-h_2)_\mu=\frac{1}{2}\left[S+L + ( L - S) \cos\theta_h\right].
  \label{eq2:Four-Body}
\end{equation}
The final pair of two-body phase spaces do not inherit the azimuthal 
symmetry, and so we do not attempt to further simplify them. We therefore 
evaluate $\mathrm{d}\Phi_2(k_2,P)$ and $\mathrm{d}\Phi_2(p_+,p_-)$ in their respective rest frames. The angles of the charged lepton frame are labelled $\theta$ and $\phi$ while those of $\Phi_2(k_2,P)$ 
are labelled $\theta'$ and $\phi'$. With these variables the four-body phase space can 
be written
\begin{equation}  \mathrm{d}\Phi_4=
  \frac{\mathrm{d}L}{2\pi}\frac{4\pi}{32\pi^2}\
 \frac{\mathrm{d}T}{S}\frac{\mathrm{d}\ell}{2\pi}\
  \overline{\beta}(k_2,P)\overline{\beta}(p_+,p_-) 
  \frac{\mathrm{d}\Omega'}{32\pi^2}\frac{\mathrm{d}\Omega}{32\pi^2}.
	\label{eq3:Four-Body}
\end{equation}
Keeping in mind that the Lorentz invariant flux factor $\mathcal{F}$ for massless initial states is given by $\mathcal{F}=2S$ we can express the parton cross section  as 
\begin{widetext}
\begin{equation}
  \sigma_{h\nu}(S)=\frac{1}{2S}\int_{m_{jk}^2}^S
  \frac{\mathrm{d}L}{2\pi}\int_L^S\frac{2\pi}{32\pi^2}\
 \frac{2~\mathrm{d}T}{S}\int_{m_{jk}^2}^L\frac{\mathrm{d}\ell}{2\pi}\
  \overline{\beta}(k_2,P)\overline{\beta}(p_+,p_-) \int \
  \frac{\mathrm{d}\Omega}{32\pi^2}\int\frac{\mathrm{d}\Omega'}{32\pi^2}\
\left|\overline{\mathcal{M} }\right|^{2}_{h\nu}.
	\label{eq4:Four-Body}
\end{equation}
\end{widetext}

\subsection{Hadron Neutrino Cross Section \label{sec:DIS-Hadron}}
We now connect our partonic cross section to the hadronic cross section
via the formalism of deep inelastic scattering. We introduce 
the new variable $S_H$ defined by $\xi S_H=S$ and is given by $S_H=2E_\nu M_N$ in the lab frame. Unlike in textbook treatments
of deep inelastic scattering, we cannot integrate $\xi$ over the full interval 
$[0,1]$ because we require a minimum amount of energy to produce the pair
of charged leptons (i.e. $\xi \approx 0$ is kinematically forbidden). Additionally we would like to ensure that  we do not double count amplitudes already included in the EPA and so we include a
cut on the minimum amount of four-momenta transfer to the nucleus 
$Q>Q_{\text{min}} $.

To impose this cut it is easiest to change from the variable $T$ to the
variable $U=Q^2=|q^2|$. If we place a cut on the momentum transfer 
$U>Q^2_{\text{min}}$ then this changes the bounds of integration in \cref{eq4:Four-Body}.
We chose $Q_\text{min}=1\text{GeV}$ to ensure we are not double-counting amplitudes. However with this scheme we include a parametric regime in which hadronic resonances can be very important.
Although the description in terms of partons may capture some of the essential
features of hadron production it is probable that the DIS formalism under-estimates 
the rates, because it does not incorporate resonance conditions. 

The effects of a cut on momentum transfer can be seen by noting that $U=S-T$, and that the bounds of 
integration require $S>T>L$. The smallest $L$, and by proxy $T$,  
can be is $m_{jk}^2$, which implies that $U=S-T<S-m_{jk}^2$. Combining 
this with the condition that $U>Q_{\text{min}}$ leads to 
$S-m_{jk}^2>Q_{\text{min}}^2$. Finally this can be converted into a 
minimum bound on $\xi$ given by
\begin{equation}
  \xi\geq\frac{ Q_{ \text{min} }^2+m_{jk}^2 }{S_H}.
  \label{eq1:Hadron-Connection}
\end{equation} 
Finally we note that depending on the magnitude and direction of the
individual leptons $Q^2$ could range from being very small, to $S-m_{jk}^2$
and so we must include the parton distribution functions inside the integral
over $U$. This leads to our final expression for the 
nucleon-neutrino cross section
\begin{widetext}
\begin{equation} \sigma_{H\nu}=\sum_h\int_{\xi_\text{min}}^1 \frac{\mathrm{d}\xi}{2\xi S_H} \
  \int_{m_{jk}^2}^{L_\text{max}}\frac{\mathrm{d}L}{2\pi}
    \int_{Q_\text{min}^2}^{\xi S_H -L}\frac{1}{8\pi}
    \frac{\mathrm{d}U}{\xi S_H}\ \int_{m_{jk}^2}^L
    \frac{\mathrm{d}\ell}{2\pi} \int\frac{\mathrm{d}\Omega}{32\pi^2}
    \frac{\mathrm{d}\Omega'}{32\pi^2}\
 \left|\overline{\mathcal{M} }\right|^{2}_{h\nu} \overline{\beta}(k_2,P)
    \overline{\beta}(p_+,p_-)\
    f^{(H)}_{h}(\xi,U)\label{eq2:Hadron-Connection}
\end{equation}
\end{widetext}
%
%
where $h$ runs over all the partons in the given nucleon $H\in\{n,p\}$ (either neutrons or protons), $L_\text{max}=\xi S_H-Q_\text{min}^2$, $\xi_{\text{min}}$ saturates the bound in \cref{eq1:Hadron-Connection} and $f^{(H)}_h(\xi,Q)$ is the parton distribution function for the 
parton $h$ in $H$. To obtain the 
neutrino-nucleus cross section a simple weighted sum of individual 
nucleon cross sections was used 
\begin{equation}
  \sigma_{A\nu}=Z\sigma_{p\nu}+(A-Z)\sigma_{n\nu}.
\end{equation}

\section{Luminosity Estimates}\label{sec:luminosity}

\subsection{SHiP}
For the purposes of calculating expected rates at SHiP we relied on 
Ref. \cite{SHiPCollaboration2015}; specifically Figure 5.25 and 
Table 2.3. These quote the number of expected charged current events in the 
detector. To convert this into a neutrino luminosity we simply divided by the 
charged current cross section which we took to be given by 

\begin{equation}
  \sigma_{CC}=A\left( \frac{E_\nu}{\text{GeV}} \right) 
  \begin{cases}
  6.75\cdot 10^{-39}~ \text{cm}^{-2} ~~~~~(\nu)\\
  3.38\cdot 10^{-39}~\text{cm}^{-2} ~~~~~ (\overline{\nu})
 \end{cases} 
\end{equation}
with the braced numbers referring to incident neutrinos and anti-neutrinos 
respectively. 
To determine the experiment's lifetime integrated luminosity, we used the number of CC events from Table 2.3
of \cite{SHiPCollaboration2015}, while the energy spectrum was taken from Figure 5.25. Finally we multiplied by the detector's efficiency (which we took to be 90\% for each of the final state leptons), leading to \cref{eq1:Prospects}.
\subsection{DUNE}
The DUNE collaboration's far and near detectors are treated separately in their
proposals, with a heavier emphasis on the far detector. As a result there is no
published neutrino spectrum for the near detector, however both detectors 
have lifetime expected event counts. We therefore had to infer the near 
detector spectrum from that of the far detector, and then normalize our results to reproduce the lifetime rates quoted  in Table 6.1 of 
\cite{DUNECollaboration2015}.

To link the beam luminosity in the far detector to those in the near detector
we also adjusted the various flavours' luminosity to account for oscillation 
effects. All $\nu_e$ appearances at the far detector were assumed to stem from
$\nu_\mu$ at the near detector, while $\nu_e+\overline{\nu}_e$ background 
in the far detector was assumed to represent the full flux of first generation
neutrinos at the near detector up to geometric losses due to beam spread.  

Additionally the CC rates in the DUNE proposal at the near detector are quoted per $10^{20}$ protons on
target (POT) and one tonne of detector mass. The far detector rates are quoted
assuming $150~\text{kt-MW-yr}$. This number assumes a $40~\text{kt}$ far 
%
%
detector, and that $1.2~\text{MW}$ of beam power corresponds to 
$1.1\times 10^{21}~\text{POT/yr}$.  We therefore multiply the event counts in 
Table 6.1 in \cite{DUNECollaboration2015} by 
\begin{equation}
  \frac{  1.1\times 10^{21}\text{POT/yr}}{1.2\text{MW}}
  \times 850\text{kt-MW-yr}\times \frac{0.1~\text{tonnes}}{40\text{kt}}
\end{equation}
where $850~\text{kt-MW-yr}$ is the exposure at the far detector in the lifetime of DUNE given the optimized design and $0.1~\text{tonnes}$ is the mass of the near detector.

Next we consider the details of the far detector. For this we use Figures 3.5, 3.29 and Table 3.5. Table 3.5 and Figure 3.5 are in correspondence with one
another, and quote their results for an exposure of $150~\text{kt-MW-yr}$. They
specify different rates for the running of the experiment in neutrino and 
anti-neutrino mode; we presume each mode constitutes half of the experiment's
lifetime. We therefore adjust the rates quoted in Table 3.5 and 
Figure 3.5 of \cite{DUNECollaboration2015} by a factor of 
\begin{equation}
  \frac{850~\text{kt-MW-yr}}{150~\text{kt-MW-yr}} \times\frac{1 }{2}
\end{equation}
to obtain the lifetime event rate for the far detector. The spectrum is
given in Figure 3.29 and here is quoted in units of CC-Events/GeV/kT/yr. The 
experiment is set to obtain an exposure of $300~\text{kt-MW-yr}$  at 
$1.07~\text{MW}$ and then $550~\text{kt-MW-yr}$  at $2.14~\text{MW}$.
Additionally the energy bin-width of the plot is $0.25~\text{GeV}$ and so we multiply the 
spectrum of Figure 3.29 of \cite{DUNECollaboration2015} by a factor of 
\begin{equation}
  \frac{0.25~\text{GeV}}{1~\text{bin}}\left( 
  \frac{300~\text{kt-MW-yr}}{1.07~\text{MW}}
  +\frac{550~\text{kt-MW-yr}}{2.14 \text{MW}} \right).
\end{equation}
Finally in Figure 3.29 the individual CC-event rates 
of $\nu_e$ and $\overline{\nu}_e$ are not given, but their sum is given.
We assumed the relative ratio of neutrinos to anti-neutrinos was equal to 
the appearance rates quoted in Table 3.5 of \cite{DUNECollaboration2015}.
%
%
Although the background neutrino rates are much smaller than the oscillation signal,
they provide the dominant contribution at the near detector. 
The production fractions of $K^+$ and $K^-$ kaons, denoted $R_{K^\pm}$ have to be compared with those of $\pi^+$ and 
$\pi^-$, given as $R_{\pi^\pm}$.

We therefore assume that at the far detector the relative components of the $\nu_e+\overline{\nu}_e$ background are given by
\begin{subequations}
\begin{equation}
N^{(\text{bkg})}_{\nu_e}= \frac{R_{\pi^+}}{R_{K^+}} \frac{N^{(\text{osc})}_{\nu_e}}{N^{(\text{osc})}_{\text{tot}}}N^{(\text{bkg})}_{\text{tot}}
\end{equation}
\begin{equation}
N^{(\text{bkg})}_{\overline{\nu}_e}= \frac{R_{\pi^-}}{R_{K^-}} \frac{N^{(\text{osc})}_{\overline{\nu}_e}}{N^{(\text{osc})}_{\text{tot}}}N^{(\text{bkg})}_{\text{tot}} .
\end{equation}
\vspace{3pt}
\end{subequations}
%
%
We then assume the first-generation component at the near detector is the progenitor of the full  background signal at the far detector. Equivalently we estimate the number of electron and anti-electron events at the near detector to be proportional to  $N_{\text{Bkg}}$ at the far detector with an overall normalization that is consistent with geometric losses. To find the geometric loss factor 
we compared the rates for $\nu_\mu$ CC events quoted in Table 6.1 of \cite{DUNECollaboration2015} with the $CC$ events from the  $\nu_\mu$ background signal  and $\nu_e$ appearance signal quoted in Table 3.5 and Figure 3.5 of \cite{DUNECollaboration2015}.   Our beam spectrum at the far detector was taken from Figures 3.29 and 3.5 of \cite{DUNECollaboration2015}.

\section{Deep Inelastic Scattering Results}

\begin{figure}[!hb]
	\centering
	\includegraphics[width=0.48\textwidth]{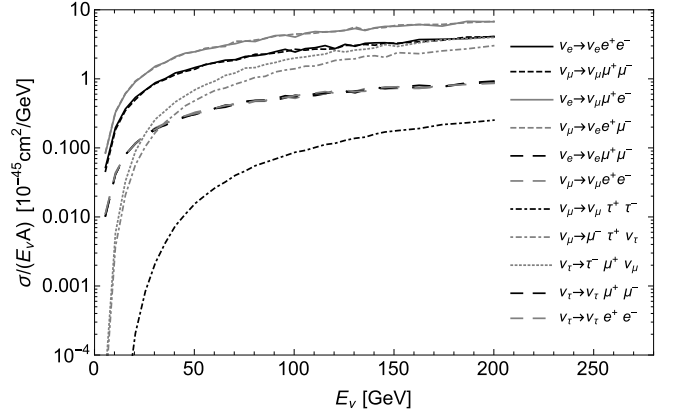}
	\caption{$\sigma/E_\nu$ trident DIS cross sections per nucleon for various SM flavours as a function of the incoming neutrino energy on a lead target (SHiP).}
	\label{fig:smsigmavseshipDIS}
\end{figure}

\begin{figure}[!hb]
	\centering
	\includegraphics[width=0.48\textwidth]{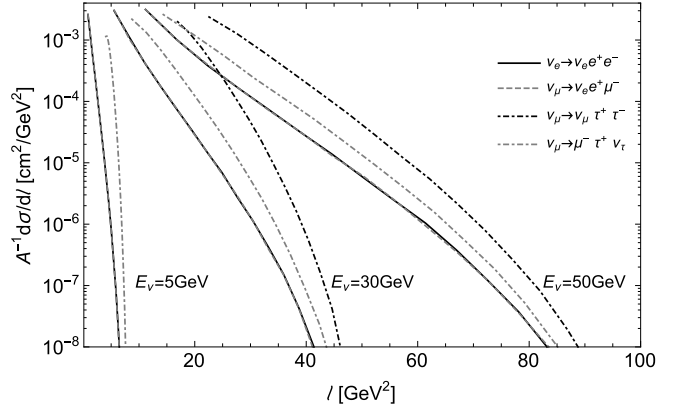}
	\label{fig:dsigmadlship}
	\caption{Normalized d$\sigma/dl$ for a variety of DIS processes at SHiP, where $l=(p_{l^+}+p_{l^-})^2$. Energies are in GeV.}
\end{figure}

\end{appendices}

\pagebreak
\bibliographystyle{abbrv}
\bibliography{trident.bib}

\end{document}